\documentclass[a4paper,11pt]{article}

\usepackage{jcappub} 

\usepackage[T1]{fontenc} 
\usepackage{aas_macros}
\usepackage{amsmath}
\usepackage{graphicx} 

\usepackage{array}
\usepackage{xcolor}
\usepackage{cancel}
\usepackage[utf8x]{inputenc}
\usepackage[left]{lineno}
\usepackage[normalem]{ulem}
\usepackage{lineno}

\title{Evolution of Neutron Star Environment in the Galactic Halo : Implications for Dark Matter Accretion} 

\author{Payaswinee Arvikar \textsuperscript{1,2}}
\author{Aseem Paranjape \textsuperscript{3}}
\author{Saee Dhawalikar \textsuperscript{4}}

\affiliation{\textsuperscript{1}Dharampeth M. P. Deo Memorial Science College, North Ambazari Road, Nagpur 440033, India}
\affiliation{\textsuperscript{2}Department of Physics, BITS-Pilani Hyderabad Campus, Hyderabad 500078, India}
\affiliation{\textsuperscript{3}Inter-University Centre for Astronomy and Astrophysics, Ganeshkhind, Post Bag 4, Pune 411007, India}
\affiliation{\textsuperscript{4}Departamento de F\'{\i}sica Fundamental, Universidad de Salamanca, E-37008 Salamanca, Spain}

\emailAdd{p20230533@hyderabad.bits-pilani.ac.in}
\emailAdd{aseem@iucaa.in}
\emailAdd{saee@usal.es}

\abstract{
Neutron stars (NS) are one of the indirect detection probes for dark matter (DM). The presence of DM is known to affect the observable properties of NS. Theoretical calculations for various equations of state of a DM admixed NS lead to estimates of the DM mass in a NS of the order of $10^{-2} M_{\odot}$. On the other hand, simplistic estimates of the amount of DM that is accreted on to the NS in the Solar neighborhood, over its age, suggest that this number is of the order $10^{-14} M_{\odot}$. Various studies have addressed this non-agreement theoretically by explaining the mechanisms leading to higher fraction than expected from smooth spherically symmetric accretion. In this work, we attempt to assess the role of the dynamic DM environmental density in the Galactic halo to explain possible enhancement in DM accretion. We consider a high resolution N-body simulation and method of Voronoi tessellation to calculate local DM density around putative NS locations in a statistically representative sample of Milky Way analogues. We infer that the dynamics of substructure may enhance the DM mass in NS by about a factor 2 as compared to the baseline, spherically symmetric expectation. Environmental effects therefore cannot explain the orders of magnitude discrepancy between equation of state and accretion based estimates of DM admixed in NS. 
}

\begin{document}
\maketitle

\section{Introduction}

The quest to uncover the nature of dark matter (DM), which constitutes the dominant fraction of the mass budget of the Universe, has motivated the development of a wide range of theoretical, experimental, and observational approaches. Significant efforts have been devoted to direct and indirect detection strategies. 
Direct detection experiments, such as XENON100 \cite{Aprile_2012} and PANDAX-II \cite{Wang_2020}, seek to observe rare interactions between DM particles and ordinary matter within highly sensitive underground detectors. Complementing these efforts, indirect searches utilize astrophysical and cosmological observations to identify signatures of DM through its gravitational effects and potential annihilation or decay products. Space-based observatories, including the James Webb Space Telescope, provide unprecedented insights into the formation and evolution of large-scale cosmic structures \cite{JWST}, offering valuable opportunities to probe the role of dark matter in shaping the Universe, together with numerous other indirect detection methods \cite{Heros2020,Biondini:2023}. 

Neutron stars (NS) being dense gravitating objects, can serve as promising sites for the accumulation of DM. Presence of DM can alter the equation of state for the NS which potentially reflects on its observable properties. To understand DM particle type, interaction, mass and its fraction in neutron stars are of interest in the current research. Constraining the fraction of DM in NS has always been an interesting and challenging aspect in such investigations.
Different studies highlight the variation in permissible DM fractions in NSs, based on the astrophysical observations. These observations include the X-ray data \cite{Riley_2021,Miller_2021, Choudhury:2024xbk} for various pulsars and the gravitational wave (GW) data \cite{Abbott_2018} for the NS merger events. Several studies have constrained the DM mass fraction in NSs using astrophysical observations and Bayesian analyses. For non-interacting DM, the inferred DM fraction is typically around $\sim$5-25\%, depending on the particle nature, equation of state, and observational constraints \cite{thakur2023exploring,Karkevandi2021,Routaray_2023,Giangrandi_2024,Ivanytskyi_2020,Shakeri:2024,Rutherford:2025}. Also considering the non-gravitational interactions, a few percent of DM in NS is inferred \cite{Shirke23}. In addition, Bayesian inference has been widely employed to constrain fermionic, bosonic, and other interacting DM models \cite{Rutherford_2023,Arvikar:2025,Rutherford:2025,Arvikar:2026,Liu_2025}. Discussions also include accumulation of asymmetric DM in compact objects \cite{Kouvaris_2010} by considering accretion of DM during all the stages of its life cycle. 

During various stages of stellar evolution, from protostar to NS, DM may get captured in the NS. The amount of DM accrued by the NS depends on the DM density, the time of exposure, the interaction cross section between the DM and baryonic matter and the type and mass of DM. In case of annihilating DM, the DM fraction may decrease and it may affect the temperature profiles. Surface temperature of a NS may fall between 3000 to 10000 K \cite{Kouvaris_2008}, as a result of annihilation process. At these temperatures, the blackbody spectrum peaks in the UV–optical band, where Galactic extinction substantially hinders accurate determinations of NS surface temperatures. Old NS or NS in denser environments may have accrued a significant amount of DM. 
Other possible mechanisms are also discussed in the literature resulting in the presence of DM in NS such as, bremsstrahlung of a DM particle \cite{Ellis_2018}, accretion of baryonic matter by the DM star \cite{Karkevandi2021} or conversion of neutron to DM explained through neutron decay anamoly \cite{Shirke23}.
In light of this, a NS closer to the Galactic center is expected to accumulate more DM mass, due to central denser regions  \cite{Ivanytskyi_2020,Del_Popolo_2020,Karkevandi2021}.
Several studies have investigated DM accumulation in neutron stars by considering different DM models and Galactic environments. Ref. \cite{Deliyergiyev:2023} estimated mass accretion from DM clumps in a Milky Way simulation, while Refs. \cite{G_ver_2014,Luo_2025} studied the accumulation of bosonic and fermionic DM, including annihilating and non-annihilating scenarios, using local DM densities, pulsar trajectories, and Navarro-Frenk-White (NFW) halo profiles to constrain DM interaction cross sections and estimate the accumulated DM mass. DM accumulated by smooth spherically symmetric accretion during the life span of the star is still not consistent with the fractions reported by various studies based on Tolman-Oppenheimer-Volkoff (TOV) \cite{OppenheimerVolkoff1939,Tolman1939} calculations, as summarized above. Theoretical NFW density \cite{Navarro1995} of halo with a spherically symmetric accretion, considers the DM halo to be static. However, one may expect the local DM density around the NS to evolve with the Milky-Way substructure over time, both systematically and stochastically. Any stochastic event of increased surrounding densities may enhance the total mass accreted over the age of NS. This suggests that the conventional assumption of a static NFW background may underestimate the total DM exposure of a NS. A more realistic treatment should incorporate the dynamical evolution of halo substructure and the resulting time-dependent density field. 

We consider a realistic situation where local density of NS is traced with the evolution of the DM halo. Cosmological simulations reveal that halos continuously evolve through mergers, tidal interactions, and ongoing accretion of matter. NS may encounter several stochastic events during its lifetime leading to spikes in the local densities. This may result in enhancements in the accumulation of mass, which can not be measured with static NFW density over the NS age as reported in earlier studies \cite{Ivanytskyi_2020,Del_Popolo_2020}. 
In this work, we study this effect of substructure motions over time on the NS environmental densities, using N-body simulations. Specifically, we exploit the statistical and dynamical reach of a recent high-resolution, cosmological simulation suite {\bf Sahyadri} (described in section \ref{methodology}), to investigate the expected distribution of DM accreted by a typical NS over its lifetime.

The paper is organized as follows. In section \ref{methodology}, we describe the N-body simulation and discuss the techniques for the densities and mass accretion calculations. In section \ref{results}, we present our results and in the last section we conclude the outcomes.

\section{Methodology}\label{methodology}

To study the effect of evolving environment of NS on DM fraction, we used part of a high resolution N-body simulation suite, \textbf{Sahyadri}, recently developed by Dhawalikar \textit{et al.} \cite{dhawalikar2026sahyadri}. 
We use the box corresponding to the default model, which is a flat $\Lambda$ cold dark matter ($\Lambda$CDM) cosmology having total matter density parameter  $\Omega_m=0.3138$, Hubble constant $H_0 = 100 $h km s$^{-1}$ Mpc$^{-1}$
with $h = 0.6736$ and primordial scalar spectral index $n_s = 0.9649$. It is performed with tree-PM code GADGET-4 \cite{Springel_2021}.
This simulation evolves $2048^3$ particles in a comoving periodic box of side length 200 $h^{-1} Mpc$, with a particle mass of $m_p = 8.1 \times 10^7 M_\odot$ with $4096^3$ PM grid and a force softening length of $\sim 3.26 h^{-1}kpc$. The resolved DM halo mass is $\sim 3.2 \times 10^9 M_\odot$ with 40 particles. The simulation generated 101 snapshots between $z = 12$ and $z = 0$, uniformly spaced in scale factor $a = (1 + z)^{-1}$ with $\Delta a = 0.01$.  Dark matter halos in the simulations were identified using the six-dimensional phase-space Friends-of-Friends algorithm implemented in ROCKSTAR \cite{Behroozi_2012rock} \footnote{https://bitbucket.org/gfcstanford/rockstar/}. Halo merger trees were constructed with the CONSISTENT TREES code \cite{Behroozi_2012trees} \footnote{https://bitbucket.org/pbehroozi/consistent-trees/}.

\subsection{Selecting MW analogs and their progenitors}

Using the simulation, we try to assess the impact of environment on DM mass accreted in NS. We start by controlling the environment at the scale of the local group (LG) of galaxies and eventually focus on halos that can host the Milky Way (MW).
We find DM haloes resembling the LG in the simulation halo catalog at redshift zero. The LG analog is selected on the basis of its mass and the major subhaloes in LG. We expect the subhaloes in LG with masses comparable to the estimated DM masses of Andromeda (M31), MW and Triangulum (M33) galaxies.
The DM halo mass of MW, M31 (\cite{Sawala_2023,Makarov_2025}) and M33 (\cite{Corbelli_2014}), were estimated and reported in various studies in the literature. The average masses are listed in the table \ref{tab:subhalo_masses}. We consider a factor 4 range of masses to select these halo analogs. We identify 255 such systems in the Sahyadri simulation box. 
We used the catalog merger trees to find the most massive progenitors of these MW analogs at the earlier time steps up to the snapshot corresponding to the age of the NS. 

\begin{table}[]
    \centering
    \renewcommand{\arraystretch}{1.5}
    \begin{tabular}{|c|c|}
        \hline
        Halo &  DM mass \\ 
        \hline
        LG & (3 $\pm$ 0.15) $\times 10^{12}M_\odot$ \\ 
        \hline
        M31 & (1.5 $\pm$ 1.4 )$\times 10^{12}M_\odot$ \\ 
        \hline
        MW & (1.1 $\pm$ 0.6) $\times 10^{12}M_\odot$ \\ 
        \hline
        M33 & (4.3 $\pm$ 0.1) $\times 10^{11}M_\odot$ \\ 
        \hline
    \end{tabular}
    \caption{Estimated DM halo masses for the galaxies.}
    \label{tab:subhalo_masses}
\end{table}

\subsection{NS placement and density evaluation} 

We place NS in the identified MW halos and, for each of them, randomly choose an age uniformly between 0.1 to 2.5 Gyr. 
The lower bound is set by the minimum time interval between consecutive simulation snapshots, while the upper bound approximately corresponds to the 99$^{th}$ percentile of the pulsar age distribution from the ATNF Pulsar Catalogue.

Considering the resolution limitation of the simulation, we put the lower limit on the reliable distance at 10 $h^{-1}$kpc ($\sim$ 3 times the force softening length).
Hence we take a conservative approach and place the NS at a distance of 20 $h^{-1}$kpc from the galactic center, with a randomly chosen direction. 
In this study, we consider two approaches for density calculation. The first is similar to the baseline approach from literature that assumes an NFW profile of host DM halo, modified to consider the profiles with evolving mass and concentration for the DM halo and its progenitors, as follows. 
\[ \rho_{NFW}(r) = \frac{\rho_0}{\frac{r}{r_s}(1+\frac{r}{r_s})^2}\] with characteristic density, \[ \rho_0 = \frac{M_{vir}}{4 \pi r_s^3 [ln(1+c)-\frac{c}{1+c}]}\] where $c=\frac{r_{vir}}{r_s}$ is the concentration, $r_s$ is the scale radius of the halo and $M_{vir}$ is the mass contained in the virial radius $r_{vir}$. We use the evolving values for $M_{vir}$ and $r_s$ from the simulated halo catalog. The NFW density calculated at the NS position is further used to calculate the mass accretion on to the NS during each small time step. 

Our new approach estimates the spatially and temporally varying density field from the simulations. To investigate the dynamical nature of the inner structure of a halo and its influence on the accretion, we used the standard method of \textit{Voronoi tessellation} 
\cite{Voronoi_1908,Paranjape_2020} to calculate DM density at the position of NS in the MW halo. 
We use a $100\ h^{-1}\text{kpc}$ sub-box centered at the halo position. For a given set of DM particles in the sub-box, called tracers, Voronoi tessellation assigns each tracer a cell with a volume. This method \cite{dhawalikar2026sahyadri} works as follows. For a given number of tracers, i.e. particle positions, a large number of random points $N_{ran}$ are generated in the selected sub-box. Each random point is assigned to the nearest tracer, using KDTrees. For each tracer the assigned number of randoms $n_{ran}$ is calculated. The volume of the associated Voronoi cell is given by, 
\[V_c = \frac{V_{total}}{N_{ran}} n_{ran}\]
Overdensity $\delta=\rho/\bar\rho -1$ in each Voronoi cell is then calculated with the known randoms in the cell and the volume of the selected sub-box. Here, $\bar\rho$ is the mean density of the sub-box and is not same as the mean density of the universe. The required density in the cell is calculated as, $\rho=(\delta+1)\bar\rho$.

In the selected sub-box NS is set at $20\ h^{-1}\text{kpc}$. Guided by a convergence study, we populated the volume with $N_{\text{ran}} = 60{,}000$ uniform random points. Scaling the box size up to $200\ h^{-1}\text{kpc}$ yielded density values consistent within $\sim 10\%$, confirming numerical stability.

We consider following two ways to place the NS in the DM halo.

\textit{Case 1 :} NS is stationary at the chosen random location at 20 $h^{-1}$kpc from the halo center. 
NS remains at this relative position for all the earlier progenitors through its age. It must be noted that halo center moves around in the simulation box through the age of the NS and the NS environment changes as the halo evolves.

\textit{Case 2 :} NS revolves around the halo center in circular orbit of radius 20 $h^{-1}$kpc. Its position and direction of orbital motion at redshift zero are chosen randomly for each halo. The mass enclosed within a sphere of radius $R = 20$ $h^{-1}$kpc is calculated for a MW halo and its progenitors. The average $M$ of these masses is used to calculate the magnitude of orbital velocity $\sqrt{GM/R}$ for that MW halo.

\begin{figure}[h]
    \centering
    \includegraphics[width=\linewidth]{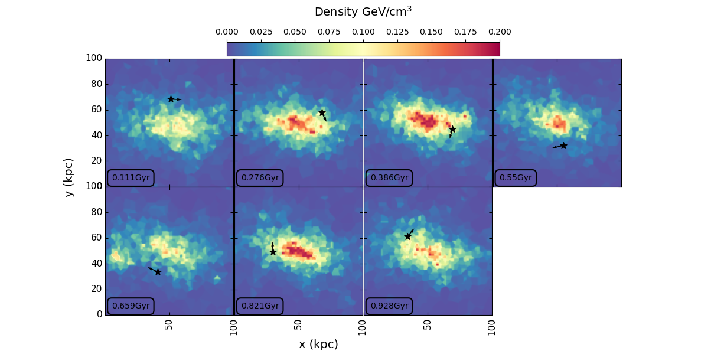}
    \caption{Black star symbol represents the NS with the arrow showing its orbital direction. Each panel showing a single-cell slice of a Voronoi tessellated density field in 100 $h^{-1}$kpc box for one MW analogue. The sub box is always placed with halo at the center. The box is sliced so as to contain the NS position at that snapshot. These are the sub boxes of the main simulation box of snapshots 100 to 96, starting from top left to bottom right. 
    }
    \label{fig:vor_NSorbit}
\end{figure}

For both the \textit{Cases} of NS placement, we calculate NFW and Voronoi densities for each MW analog. Figure \ref{fig:vor_NSorbit} shows the NS positions on the Voronoi tessellation plots in seven snapshots. It shows the density of the particle distribution in $GeV/cm^3$.For representational purposes, Voronoi densities are interpolated onto a $128^3$ grid in the 100 $h^{-1}$kpc sub-box. This plot corresponds to the most massive of the MW analogs found, of virial radius 216.3 $h^{-1}$kpc and mass $1.21 \times 10^{12}$ $M_\odot h^{-1}$. Its chosen age is 0.88 Gyr. The arrow with the NS position indicates the direction of the circular orbit. It can be seen from figure \ref{fig:vor_NSorbit} that NS being at the same distance from halo center encounters different densities at different epochs. We further compare the mass accretion with NFW densities and Voronoi densities for the two \textit{Cases}, stationary and orbiting NS.

\subsection{DM mass accretion}

In neutron star astrophysics, TOV based calculations expect a few percent of DM in NS, reported in various studies considering different DM models. On the other hand, the reported accreted DM mass is orders of magnitude smaller, considering a NS at fixed positions over its age and its non-evolving environment over time. Here we attempt to investigate the non-agreement between DM mass in NS, based on DM accretion and the TOV estimates, considering the possible local density variations. 

NS over its age, may change its position with respect to the halo center and the local DM density also evolves with host halo. We consider $\rho_\chi$ as a spatially local, time-averaged quantity that follows the NS position: $\langle\rho_\chi\rangle = \frac{1}{t}\int_0^t{\rm dt^\prime}\rho_\chi(\mathbf{x}(t^\prime),t^\prime)$.
As a first simplistic approximation, we consider NS fixed at a desired distance from halo center, i.e., $\mathbf{x}$ does not change with time $t^\prime$, which is stated as \textit{Case 1}. Our \textit{Case 2} represents the second approximation, where $\rho_\chi$ is a function of NS position and time ($\mathbf{x}(t^\prime),t^\prime$).  

The standard approach in the literature assumes NS at a fixed distance and local density. For a typical NS of mass $M = 1.4M_{\odot}$, radius $R = 10$ km, and for non-annihilating DM, the accreted mass is given by, \cite{Kouvaris:2013awa}
\begin{equation}{\label{Macc}}
    M_{acc} \approx 10^{-14} \bigg(\frac{\langle\rho_{\chi}\rangle}{0.3 {GeV/cm^3}}\bigg) \bigg(\frac{\sigma_{\chi n}}{10^{-45}cm^2}\bigg) \bigg(\frac{t}{Gyr}\bigg) M_\odot \,.
\end{equation}
The cross-section $\sigma_{\chi n} \sim 10^{-45} cm^2$ for DM particle mass of some hundreds of GeV \cite{XENON_2017}. 

In this study, we calculate $\langle\rho_\chi\rangle \times t$ integrated over the NS age for both the \textit{Cases}. It is calculated for each MW analog considering NFW as well as local densities with Voronoi tessellation, through each small time step of the simulation. With this value in Eq. \ref{Macc} we calculate total $M_{acc}$ in both the \textit{Cases}.

\section{Results and Discussions}\label{results}

This work aims to investigate whether the DM mass accumulated by NS can be significantly enhanced beyond the predictions of the static NFW halo model. Specifically, we examine whether the time evolution of the Galactic DM halo, as captured by cosmological N-body simulations, can account for greater DM accretion than through spherically symmetric capture. Furthermore, we explore whether encounters with DM substructures along the trajectories of NSs over their lifetimes lead to enhanced DM accretion compared to that predicted using the smooth NFW density profile.

For the NFW densities, i.e. the spherically symmetric mass distribution, the accumulated mass in a NS placed at 20 $h^{-1}$kpc is of the order of $10^{-15} M_\odot$. This value depends on distance and age, hence can increase to about $10^{-10} M_{\odot}$ if NS is around 0.1 $h^{-1}$kpc and of age 10 Gyr. Based on the analysis above, we estimate the additional impact of an evolving halo environment on the accreted DM in the NS, in the two Cases described earlier.

\begin{figure}[h]
    \centering
    \includegraphics[width=\linewidth]{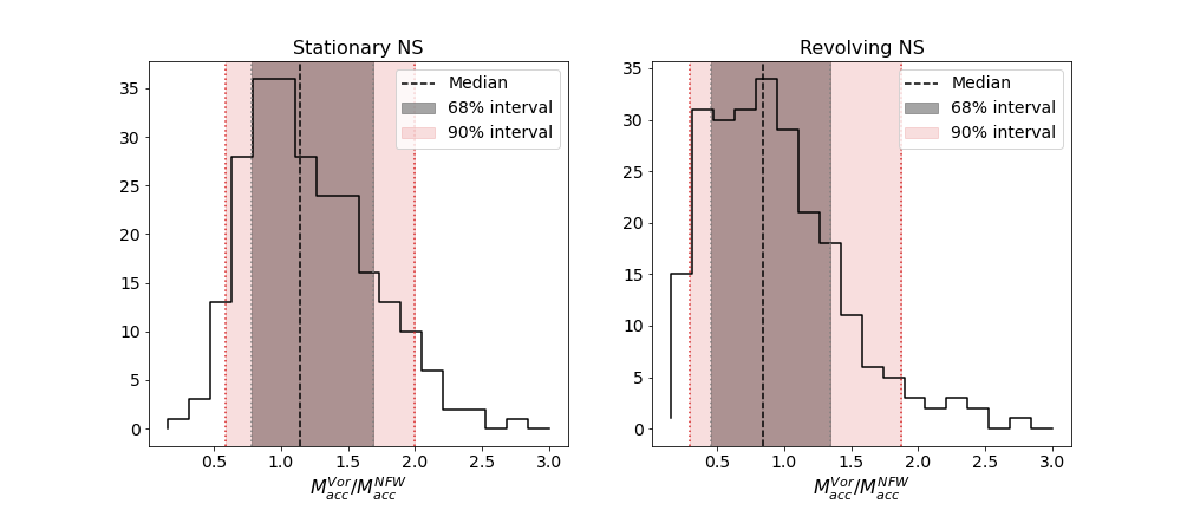}
    \caption{Distribution of the ratio of DM mass accretion using Voronoi density to NFW density. \textit{Case 1} for stationary NS (Left panel) and \textit{Case 2} for revolving NS (Right panel). Black dashed line shows the median of the data and the gray and red shaded regions correspond to the 68\% and 90\% confidence intervals respectively.}
    \label{fig:Macc_distri}
\end{figure}

Figure \ref{fig:Macc_distri} shows the distribution of the ratio $M_{acc}^{Vor} / M_{acc}^{NFW}$ of DM mass accretion on to the NS placed at 20 $h^{-1}$kpc ie. \textit{Case 1} (Left panel) and for a NS revolving in a orbit of radius 20 $h^{-1}$kpc ie. \textit{Case 2} (Right panel), with the value expected in the baseline, NFW case. 
Even for a stationary NS the local DM environment changes as the halo evolves with time. Hence total mass accretion with NFW density may differ from that using Voronoi density for the same age. This can be seen in the figure \ref{fig:Macc_distri} (Left panel), where $M_{acc}^{Vor} / M_{acc}^{NFW}$ has a larger spread on values greater than 1 than on lower values, although the median is at 1.14. Also for revolving NS (right panel) the ratio has median at 0.81 but a longer tail towards 2.5. The shaded regions correspond to the confidence intervals of 68\% (gray) and 90\% (red). Considering the 95$^{th}$ percentile of the data shows that mass accretion can be enhanced by a factor $\sim2$ when we consider local densities over NFW densities.\footnote{To check the effect of distance from the halo center, we repeated the analysis for a stationary NS at 10$h^{-1}kpc$. The distribution of $M_{acc}^{Vor} / M_{acc}^{NFW}$ for this analysis also shows at most factor 2 enhancement.}
Hence, it can be inferred that an order-of-magnitude enhancement in the mass accretion is not expected due to environmental effects. In addition, we cannot explain the mass of the DM of around 5\%, as predicted by the TOV calculations on the basis of the dynamics of the substructure in the DM halo. 
Instead, as mentioned in the Introduction, one may need to invoke other mechanisms like neutron decay anomaly, which may lead to conversion of neutrons to DM particles in NS, or accumulation of DM during the main sequence and supernova phases of the NS.

\section{Conclusion}\label{conclusions}

We have performed a realistic assessment of the impact of environmental effects in the DM mass accreted by a typical NS over its age. It takes into account the evolution of the DM halo where the MW resides in and the possible motion of NS. The results indicate at most factor $\sim 2$ enhancement, at $95\%$ confidence, in accreted DM mass. Hence we can conclude that, the environmental effects are not expected to change the order of magnitude of accreted mass, compared to simplistic estimates based on smoothly varying NFW profiles. The DM mass fractions based on TOV calculations can not be explained only on the basis of substructure evolution in the very dense regions of the galactic halo.

\section*{Acknowledgments}
We gratefully acknowledge the use of high performance computing facilities at IUCAA, Pune. PA thanks IUCAA for the research facilities. PA also thanks Sarmistha Banik for the insightful discussions.

\bibliographystyle{JHEP} 
\bibliography{DM_NS}
\end{document}